\documentclass[a4paper,12pt]{article}
\usepackage{latexsym,amsmath}
\usepackage{amssymb}
\usepackage[dvips]{graphicx}
\usepackage{lscape}
\usepackage{cite}
\usepackage{color}
\usepackage{accents}
\usepackage[T1]{fontenc}
\usepackage[utf8]{inputenc}
\usepackage{authblk}
\usepackage[authoryear,round,longnamesfirst]{natbib}
\usepackage{booktabs}
\usepackage{multirow}
\usepackage{siunitx}
\usepackage{float}
\usepackage{setspace}
\usepackage{subcaption}
\usepackage{tcolorbox}
\usepackage{mathtools}
\usepackage{epstopdf}
\usepackage{geometry}
\usepackage{multicol}
\usepackage{multirow}
\usepackage{csquotes}
\usepackage{hyperref}
\hypersetup{
     colorlinks   = true,
     linkcolor = blue,
     anchorcolor = blue,
     citecolor = blue,
     filecolor = blue,
     urlcolor = blue
}

\usepackage{etoolbox}
\usepackage{rotating}
\usepackage{tikz}
\usepackage{enumitem}   
\usetikzlibrary{decorations.pathreplacing}
\usepackage{lineno}
\usepackage{setspace}
\usepackage{hyperref}
\usepackage{mathrsfs}
\usepackage{bigints}
\usepackage{amsthm}
\usepackage{fancyhdr}
\newtheorem{lemma}{Lemma}

\newtheorem{theorem}{Theorem}

\patchcmd{\abstract}{\small}{}{}{}

\newcommand{\uiota}             {\mbox{\boldmath$\uiota$}}

\def\beq{\begin{equation}}
\def\eeq{\end{equation}}
\def\beqa{\begin{eqnarray}}
\def\eeqa{\end{eqnarray}}
\def\beqan{\begin{eqnarray*}}
\def\eeqan{\end{eqnarray*}}

\def\bc{\begin{center}}
\def\ec{\end{center}}
\def\btable{\begin{table}[htbp]}
\def\etable{\end{table}}
\def\bfig{\begin{figure}[htbp]}
\def\efig{\end{figure}}

\def\bi{\begin{itemize}}
\def\ei{\end{itemize}}

\let\oldabstract\abstract
\let\oldendabstract\endabstract
\makeatletter
\renewenvironment{abstract}
{%
               {\list{}{\addtolength{\leftmargin}{1em} 
                        \listparindent 1.5em%
                        \itemindent    \listparindent%
                        \rightmargin   \leftmargin%
                        \parsep        \z@ \@plus\p@}%
                \item\relax}%
               {\endlist}%
\oldabstract}
{\oldendabstract}
\renewcommand{\@makefnmark}{\hbox{\textsuperscript{\tiny{\@thefnmark}}}}
\makeatother

\title{
{\LARGE \bfseries Bayesian Variable Selection for Multi-Outcome Models Through Shared Shrinkage}\\
}
\author{{Debamita Kundu, Riten Mitra, Jeremy T. Gaskins}\\
\textit{ \href{mailto:debamita.kundu@louisville.edu; ritendranath.mitra@louisville.edu; jeremy.gaskins@louisville.edu}{\small{debamita.kundu@louisville.edu; ritendranath.mitra@louisville.edu; jeremy.gaskins@louisville.edu} }}}
\affil{Department of Bioinformatics and Biostatistics, University of Louisville, Louisville, KY 40202, USA}
\date{}
\providecommand{\keywords}[1]{\textbf{\textit{Keywords:}} #1}

\begin{document}

\maketitle
\begin{abstract}
 Variable selection over a potentially large set of covariates in a linear model is quite popular. In the Bayesian context, common prior choices can lead to a posterior expectation of the regression coefficients that is a sparse (or nearly sparse) vector with a few non-zero components, those covariates that are most important. This article extends the “global-local” shrinkage idea to a scenario where one wishes to model multiple response variables simultaneously. Here, we have developed a variable selection method for a $K$-outcome model (multivariate regression) that identifies the most important covariates across all outcomes. The prior for all regression coefficients is a mean zero normal with coefficient-specific variance term that consists of a predictor-specific factor (shared local shrinkage parameter) and a model-specific factor (global shrinkage term) that differs in each model. The performance of our modeling approach is evaluated through simulation studies and a data example.
\end{abstract}

\keywords{Global-local shrinkage prior, Multi-outcome model, Multivariate regression, Shrinkage, Variable selection}

\section{Introduction}

In the context of high-dimensional data, it is critical to correctly identify  a set of variables that significantly influences the responses and play an important role in prediction. Consider a set of $p$ potential regressors $X_1, X_2, \ldots, X_p$ and a single response variable $Y$. In order to increase the precision of statistical estimates and prediction, we often consider a model of the form
\[ Y=\beta_0+X_1\beta_1+ X_2\beta_2+ \ldots+ X_p\beta_p+\epsilon,\]
where many of the $\beta$ are exactly zero, so that only the set of $q \left(\leq p \right)$ regressors impact the response $Y$.

In the Bayesian context there are numerous approaches to the problem of variable selection. \cite{mitchell1988bayesian} proposed the \enquote{spike and slab} approach by considering a mixture prior distribution for the regressor coefficient: a zero component (spike) and a disperse component (slab). Specifically, indicator variables were used to differentiate the important regressors from the rest. When the indicator assumes the value 0, the prior for the corresponding regressor coefficient is set to follow a Gaussian with low variance. This is the zero  component (spike). Otherwise it follows a Gaussian with high variance, representing the disperse component (slab). For this setup, \cite{GeorgeMc} suggested stochastic search variable selection (SSVS) for identifying a \enquote{promising} subset. This framework was later extended to incorporate several non-conjugate and conjugate priors for prior specification \citep{george1997approaches}. Subsequently, a related class of variable selection priors that put positive mass at 0 are based on Reversible Jump (RJ) sampling techniques \citep{green2009reversible}. However, these selection methods require updating each regression coefficient conditionally on all others and tend to be computationally slow and display poor mixing when used for a large number of variables. 

Hence, shrinkage priors have gained popularity recently as a computationally faster alternative. Rather than using a mixture prior that can set the coefficient exactly to zero, the shrinkage approach employs priors designed to pull small signals aggressively towards zero. Many of the commonly used shrinkage models fall within the global-local (GL) shrinkage framework defined by \cite{polson2010shrink}. In the usual multiple regression setting where the regression coefficient vector $\beta=\left(\beta_1,\beta_2,\ldots,\beta_p \right) $
is believed to be sparse, the typical GL shrinkage prior for the $\beta$ vector would be
\[\beta_j  \sim  N (0,\lambda_j^2\tau^2),\]
\[\lambda_{j}  \sim f(\cdot), \quad \quad \tau  \sim  g(\cdot). \]
In this model $\tau$ controls global shrinkage towards the origin, and $\lambda=\left(\lambda_1,\lambda_2, \cdots, \lambda_p \right)$ are the local shrinkage parameters that allows deviation in the degree of shrinkage between predictors. The typical recommendation is that $f(\cdot)$ should have heavy tails to avoid over-shrinking large signals, and $g(\cdot)$ should have substantial mass near zero. The Normal-gamma prior \citep{griffin2010}, the Dirichlet-Laplace prior \citep{Bhatt} and the horseshoe prior \citep{carvalho2010horse} are three popular methods in this framework. A review and comparison of various variable selection methods including the shrinkage methods can be found in \cite{o2009review}.

Although much of the literature focuses on the situation of multiple regression with a single response variable $y$, the problem of variable selection when simultaneously analyzing multiple responses (multivariate regression) is much less explored. For example, multiple outcomes measuring different aspects of a patient's health (blood pressure, glucose, etc.) may be modeled using a potentially large set of risk factor predictors. In many cases, each outcome is analyzed separately with variable selection performed unique to each outcome, but this will be inefficient if each model has the same or similar set of relevant predictors. However, borrowing strength across regression coefficients can boost the power of detecting true signals, especially if the responses share similar predictors and there is reason to believe that they exert similar influences on the responses. The gain in performance can be substantial for low to moderate sample sizes and complex noise structures. Instead of applying variable selection separately for each outcome, \cite{brown1998multivariate,brown1999choice} propose two approaches based on finding a common set of predictors for all models by extending the \citeauthor{GeorgeMc}'s selection model (1993; 1997).  However, by requiring predictors to affect either all $K$ outcomes or none of them, their models are often overly restrictive. Hence, in this work we focus on developing a more flexible variable selection method that encourages the inclusion of similar sets of predictors in each of the $K$ models by extending the GL shrinkage framework. Recently, \cite{bai2018high} independently explored a similar setup and proposed their Mutlivariate Bayesian Model with Shrinkage Priors (MBSP).  We will discuss differences that distinguish our work in later sections.  In a frequentist setting, \cite{turlach2005simultaneous} proposed a LASSO-based approach with penalties based on the maximum absolute coefficient across all outcomes for each predictor.

The layout of this manuscript is as follows. In section \ref{section 2}, we describe a general strategy for GL shrinkage in multivariate regression. and explore details when paired with the 3 common GL models, Normal-gamma, Dirichlet-Laplace, and horseshoe, as well as relevant posterior consistency results. Section \ref{section 3} discusses posterior sampling for each of these models, and Section \ref{section 4} considers simulation studies to explore the performance of our model.  In Section \ref{section 5} we analyze a real data set based on the yeast cell cycle data \citep{chun2010sparse}, and we conclude with a brief discussion in Section \ref{section 6}. 

\section{Multi-outcome Regression Coefficient Shrinkage Model} \label{section 2}
\subsection{General Strategy}

Consider a multi-outcome (multivariate) model with $K$ outcomes/responses, $p$ covariates and $n$ independent observations. We write the multivariate regression model in the following form, 
\begin{equation} 
\begin{bmatrix}
    y_{11}       & y_{12} &  \dots & y_{1K} \\
    y_{21}       & y_{22} &  \dots & y_{2K} \\
    \vdots & \vdots &  \ddots & \vdots \\
    y_{n1}       & y_{n2} &  \dots & y_{nK}
\end{bmatrix}
=
\begin{bmatrix}
    x_{11} & x_{12} &  \dots  & x_{1p} \\
    x_{21} & x_{22} &  \dots  & x_{2p} \\
    \vdots & \vdots &  \ddots & \vdots \\
    x_{n1} & x_{n2} &  \dots  & x_{np}

\end{bmatrix}
\begin{bmatrix}
    \beta_{11} & \beta_{12} &  \dots  & \beta_{1K} \\
    \beta_{21} & \beta_{22} &  \dots  & \beta_{2K} \\
    \vdots & \vdots &  \ddots & \vdots \\
    \beta_{p1} & \beta_{p2} &  \dots  & \beta_{pK}

\end{bmatrix}
+ \varepsilon,
\label{multioutcome_model}
\end{equation} 
where $Y_{i\cdot}$, the $i^{th}$ row of the $n\times k$ matrix $Y$, consists of the $K$ responses for the $i$th observation and $X_{i\cdot}$ is the $i^{th}$ row of the model matrix $X$ which contains the $p$ predictor variables for this observation. The matrix of regression coefficients $B$ is believed to be sparse.  Further, as each row of $B$ corresponds to the regression coefficients of predictor $j$ on each of the $K$ responses, we expect similar sparsity across the row. $\varepsilon$ is the $n\times K$ residual matrix. Under the normality assumption, each row of the residual matrix follows a $N_K(0,\Psi)$ distribution independently. For simplicity, we ignore the intercept terms for right now. Note also that throughout we assume that the columns of $Y$ and $X$ have been standardized. This gives a multivariate normal distribution for the vector of responses for patient $i$, $ Y_{i\cdot} \sim MVN_{K} \left( X_{i.}B,\Psi\right)$. 


Variable selection is induced through the choice of prior on the $B$ matrix.  Our approach is to extend the global-local shrinkage framework to jointly model multiple responses.  The general idea of our method is to share information about the importance of a covariate across all response models through a local-shrinkage parameter $\lambda=\left(\lambda_1,\lambda_2,\hdots,\lambda_p \right)$ and use a response-specific global shrinkage parameter $\tau=\left( \tau_1,\tau_2,\hdots,\tau_K \right)$ to allow for different scalings of the regression coefficients in the different response models. Following the usual GL framework, our prior for the coefficient matrix $B$ comes from the following general hierarchy,
\begin{align}
  \begin{aligned}
   \beta_{jk}  & \sim N(0, \lambda_j^2\tau^2_k), \quad \left(j=1,2,\cdots, p, \quad k=1, 2, \cdots, K \right),\\
 \lambda_j & \sim f(\cdot),  \\
   \tau_k & \sim g(\cdot). 
  \end{aligned} \label{eq:1}
 \end{align}
 
The choices of the local distribution $f(\cdot)$ and the global distribution $g(\cdot)$ can be borrowed from any of the common global-local models. In particular, we focus on the utility of this approach under the following three choices: the Normal-gamma prior \citep{griffin2010}, the horseshoe prior \citep{carvalho2010horse}, and the Dirichlet-Laplace prior \citep{Bhatt}. The value of the local parameter $\lambda_j$ will encourage similar levels of shrinkage/sparsity for all coefficients $\left(\beta_{j1}, \beta_{j2}, \cdots, \beta_{jK} \right)$ of the $j^{th}$ predictor. Following the usual GL shrinkage rules, we choose the local distribution $f(\cdot)$ to have heavy tails and $g(\cdot)$ to have substantial mass near zero \citep{polson2010shrink}. A large $\lambda_j$ allows $\beta_{jk}\left( k=1, 2, \ldots, K\right) $ coefficients far from zero, whereas a small $\lambda_j$ will ensure all coefficients for predictor $j$ are aggresively shrunk toward zero. Note that if there is only a single response $K=1$, then our approach is exactly equivalent to the usual global-local framework. Finally, note that the general framework $\left(\ref{eq:1}\right)$ specifies the distributions $f(\cdot)$ and $g(\cdot)$ for the global and local parameters on the scales of the standard deviation of $\beta_{jk}$. In some cases, it may be more natural for $f(\cdot)$ and/or $g(\cdot)$ to represent the distribution for the variance contributions $\lambda_j^2$ and $\tau^2_k$, respectively.

Despite similarities of our framework to that of \cite{bai2018high}, there are several key differences between our approaches. First, their MBSP model specifies a common value $\tau$ for the global $\tau_k$ parameters across all models. Further, this parameter is a priori fixed based on asymptotic arguments. Conversely, we recognize that there may be variability in the global scale of the coefficients between response models, and we allows differing $\tau_k$ which are estimated from the data. Secondly, MBSP specifies the column covariance of $B$ to be proportional to $\Psi$, the residual covariance matrix.  This choice facilitates additional conjugacy in their sampler, but we opt to allow the columns of $B$ to be independent (given the $\tau_k$s) as a more intuitive choice.  As will be shown in Section 3, we are able to retain a high degree of conjugacy and develop an efficient posterior sampler.

Having defined our general approach, we now focus on three versions of our methodology by using common shrinkage models.

\subsection{Multi-outcome Normal-gamma Model }

First, we apply the Normal-gamma shrinkage prior from \cite{griffin2010} to our method. We refer this model as the Multi-outcome Normal-gamma Model $\left(\text{MONG} \right)$. This yields the following hierarchy:
\begin{align}
\begin{aligned}
\beta_{jk} & \sim N (0,\lambda_j \tau^2_k), \quad \quad \left(j=1, 2,\ldots, p ; \quad k=1, 2, \ldots, K \right), \\
\lambda_{j} & \sim  Gamma\left( c,\frac{1}{c}\right),  \\
\tau_k & \sim  C^+\left( \gamma \right),  \\
c & \sim  Exp(\lambda_c).  \label{NG prior}
\end{aligned}
\end{align}

In $\left(\ref{NG prior}\right)$, $\lambda_j$ comes from a $Gamma \left( c, \frac{1}{c} \right)$ distribution such that the prior mean is 1 and variance is $\frac{1}{c}$. Hence, small values of $c$ will induce greater variability within the $\lambda$s and more shrinkage. The tail of $\beta_{jk}$ thickens with increasing $c$. A common special case involves setting $c=1$ which provides the Bayesian LASSO \citep{park2008bayesian}. For the prior distribution of $\tau$, we consider a half-Cauchy distribution with density $f(x)=\frac{2\gamma}{\pi \left( \gamma^2+x^2\right) }, x >0$. The intuition behind considering half-Cauchy prior for global shrinkage parameter is its non-zero density near the origin with thick tails in the extremes. We recommend setting the scale parameter of this half-Cauchy to $\gamma=0.5$ to provide a reasonably dispersed distribution for the $\tau$s, and this choice has performed well in empirical studies. For the hyper-parameter $c$ we consider an exponential density with mean 2 to encourage slightly thicker tails in $\beta_{jk}$ than the Bayesian LASSO.

\subsection{Multi-outcome Horseshoe Model }

The horseshoe prior is one of the most appealing and commonly used shrinkage priors in the literature. It became popular due to its infinitely tall spike in the density near the origin that shrinks almost everything towards zero and its flat, Cauchy-like tails that allow some parameters to escape from shrinkage. The conventional horseshoe prior places half-Cauchy priors on both the local and global contributions to the standard deviation. The Multi-outcome Horseshoe Model $\left(\text{MOHS} \right)$ is defined by the following hierarchy: 
\begin{align}
\begin{aligned}
\beta_{jk}   \sim N (0,\lambda_j^2 \tau_k^2),  \\
\lambda_{j}  \sim  C^+\left( 1\right),  \\
\tau_k \sim  C^+\left( 1 \right). \label{Con_Horse prior}
\end{aligned}
\end{align}

 In its usual form, the model $\left(\ref{Con_Horse prior}\right)$ is not conjugate, making implementation in a standard Gibbs sampling scheme difficult and time-consuming. However, \cite{makalic2016simple} proposed an efficient, conditionally conjugate sampling algorithm for fast updating by introducing data augmentation variables from an inverse gamma distribution. 
Since the marginal distribution of $\chi$ from the hierarchy $\chi^2 \mid \Upsilon \sim IG\left( \frac{1}{2},\frac{1}{\Upsilon}\right)$ and $\Upsilon \sim IG\left( \frac{1}{2},1 \right)$ is $C^+\left(1\right)$, we equivalently write this model as
\begin{align}
\begin{aligned}
\beta_{jk} \sim N\left(0, \lambda_j^2\tau^2_k \right), \\
\lambda^2_j \sim IG\left( \frac{1}{2},\frac{1}{\nu_j}\right), \\
\tau^2_k \sim IG\left( \frac{1}{2},\frac{1}{\omega_k}\right), \\
\nu_1, \nu_2, \ldots, \nu_p, \omega_1, \omega_2, \ldots, \omega_K \sim IG\left(\frac{1}{2},1 \right). \label{Horse prior}
\end{aligned}
\end{align}
Note that we define $IG$ to have density function $f(x\mid \alpha,\beta)=\frac{\beta^\alpha}{\Gamma{\left(\alpha\right)}}x^{-\alpha-1}e^{-\frac{\beta}{x}}, \, x>0$. 

In both the MONG and MOHS versions, we may use the $\lambda$ parameters to judge the importance of a predictor across all responses. The larger the local parameter the less shrinkage in the regression coefficients and the greater the predictive power. Hence, the estimated $\hat{\lambda}_j$ can be used as a summary of the importance of predictor $j$ across all $K$ models.  In both cases, we may compare this value relative to 1, the prior mean for $\lambda_j$ in MONG and the prior median in MOHS.

\subsection{Multi-outcome Dirichlet-Laplace Model }
In a similar manner, we also define the Multi-outcome Dirichlet-Laplace Model $\left(\text{MODL} \right)$. Like the previous GL methods, the DL model considers the dispersion of the $j^{th}$ coefficient to be a contribution of local and global scaling terms. However, the conditional distribution of the coefficient  is Laplace (double exponential) instead of the usual normal distribution. While this may not technically fall in \cite{polson2010shrink}'s GL framework, it is clearly in the same spirit, and can be paired with our multi-outcome shrinkage framework. The proposed MODL model has the following specification
\begin{align}
\begin{aligned}
\beta_{jk} & \sim DE\left( \phi_j\tau_k\right), \\
\tau_k &\sim  Gamma\left( pa,\frac{1}{2}\right), \\
\phi=\left( \phi_1,\phi_2,\ldots,\phi_p\right)  & \sim  Dirichlet\left( a,a,\hdots,a \right),  \label{Dir prior}
\end{aligned}
\end{align}
where $a$ is concentration parameter of the Dirichlet distribution. In this model the local parameters $\phi_j$ sum to one, and smaller values of $a$ will lead $\phi$ to be dominated by a few components. Since the majority of the DE scales $\phi_j\tau_k$ will be approximately zero, sparsity in the $\beta_{jk}$ is achieved. As recommended by \cite{Bhatt}, we considered $a=\frac{1}{2}$ or $a=\frac{1}{p}$ for our simulation and case study. 

Similar to the HS model $\left(\ref{Horse prior}\right)$, we can introduce auxiliary variables to facilitate sampling. One may represent the $\beta_{jk}  \sim DE\left( \phi_j\tau_k\right)$ as scale mixture of normals through $\beta_{jk} \mid \eta_{jk} \sim  N\left(0, \eta_{jk}\phi_j^2\tau_k^2\right)$ with $\eta_{jk} \sim Exp\left( \frac{1}{2}\right)$. Similar to using $\hat{\lambda}$ to evaluate predictor relevance in the MONG and MOHS models, in this MODL proposal we can compare the estimated $\phi_j$s to their prior mean $1/p$.  Again, larger values indicate less shrinkage and greater predictor relevance across all outcomes.

Across all models for the regression coefficients the residual covariance matrix is given an inverse Wishart prior with $K+2$ degrees of freedom and the identity matrix as the prior scale matrix. This gives the prior mean for $\Psi$ as the identity matrix. As is common, we recommend responses and predictors be centered and scaled prior to analysis.

\subsection{Posterior Consistency}

In this section, we present a result guaranteeing posterior consistency in our model structure. For this proof, we will assume that the residual covariance matrix $\Psi$ is fixed and known. We  first state the assumptions before proving our consistency result. 
 
 \noindent {\bf Assumptions:} 
 \begin{enumerate}
	 \item[(A1)] The prior $\pi(B)$  is continuous in $B$ over all of $\mathbb{R}^{p\times K}$.
	\item[(A2)] The vector of covariates are uniformly bounded.  That is, there exist $G>0$ such that $||X_{i\cdot}|| < G$ for all $i=1,\ldots,n$.
		\item[(A3)] The smallest eigenvalue of the design matrix is asymptotically bounded away from zero.  There exists $c>0$ such that $\liminf_{n \rightarrow \infty} \lambda_1( \frac{1}{n}X^{\prime} X ) >c$, where $\lambda_1(M)$ refers to the smallest eigenvalue of the matrix $M$.
 \end{enumerate}
 
Note that (A1) represents a much more general class of prior models than our GL shrinkage framework, although our proposal clearly falls within this assumption. Throughout, we use the Frobenius norm, $||M||=\sqrt{\sum_{i,j} (m_{ij})^2}$.  Note also that any deterministic functions of the $n \times p$ design  matrix $X$ depends on the sample size $n$.  To avoid cumbersome notation we typically suppress the dependence on $n$ and refer to it as simply $X$.


First, we state our key theorem about posterior consistency. 
\begin{theorem} 
Assume a fixed, positive definite $\Psi$ and assumptions (A1)-(A3).  Let $Y_{1\cdot},\ldots,Y_{n\cdot}$ be iid from model $\left(\ref{multioutcome_model}\right)$ under the true parameter value $B_0$.  Then for any $\epsilon>0$,
\[P_{B_0}\left\{ \,||B- B_0||>\epsilon \mid Y_{1\cdot},\ldots,Y_{n\cdot}\right\} \rightarrow 0, \hspace{1em} \text{a.s.\ as $n\rightarrow \infty$.}\]  
\end{theorem}
That is, the posterior distribution for $B$ almost surely collapses to the true value $B_0$ as $n\rightarrow\infty$.

This proof along with the associated lemmas  appears in the Appendix. It builds upon Schwartz's  seminal proof \citep{scr}, in combination with results for regression models from \cite{ame} and \cite{choir}. The argument mainly relies on the existence of an uniformly exponentially consistent (UEC) sequence of tests and a prior positivity property. The latter in Schwartz's original proof was  simply the condition that the prior mass on all Kullback--Leibler (KL) neighborhoods of the true parameter is greater than zero. However, as we show in the Appendix, this KL framework must be modified into a  multi-index version for its use in models with covariates. Both of these two conditions are derived as separate lemmas that can be combined to give posterior consistency.  See the Appendix for full details.
  
An important feature of Theorem 1 is its flexible prior condition stated in (A1). This relaxation comes at a cost, mainly assumption (A2), which essentially bounds the  entries of the design matrix. In contrast, \cite{bai2018high} assume upper and lower (asymptotic) bounds on the eigenvalues of the design matrix. However, the flexibility gained under our choice is significant, as we require no condition (except continuity) on the prior for $B$. This is much more general than the assumptions made in the consistency theorems of \cite{bai2018high} and \cite{armr}.  Their choices require conditions on the prior with convoluted formulas involving $\Psi$ and the eigenvalues of the design matrix, thus restricting the choice of prior on $B$ in ways that are not straightforward.


\section{Posterior Computation}\label{section 3}
As with most modern Bayesian models, inference is performed by approximating the posterior through Markov chain Monte Carlo (MCMC) methods.  We describe the necessary sampling steps for each of our three models below.

\subsection{MONG Model}
\begin{enumerate}[label=(\roman*)]
\item Sample $vec( B)\mid X,\Psi,\lambda,\tau$ from $MVN_{pK}\left( M,W\right)$, where $W= \left( \left( \Psi^{-1}  \otimes X^T X \right)+\Omega^{-1}\right)^{-1}$ and 
$M = W\left(  \Psi^{-1} \otimes X^T \right) vec(Y) $. Here, $\Omega=T \otimes \Lambda$ the prior covariance matrix of $vec(B)$, $\Lambda=diag(\lambda_1, \lambda_2, \cdots, \lambda_p)$ and $T=diag(\tau_1^2, \tau_2^2, \cdots, \tau_K^2)$. Throughout, we let $\otimes$ denote the Kronecker product.
\item For $j=1,2, \cdots, p$, sample $\lambda_j \mid \beta_{jk},\tau_k,c \sim giG\left( c-\frac{K}{2},2c,\sum_{k=1}^{K} \frac{\beta^2_{jk}}{\tau^{2}_k } \right)$, where $giG\left(\kappa, \chi, \rho \right) $ is the generalized inverse Gaussian distribution with density
$f\left( x; \kappa,\chi, \rho \right) \propto x^{\rho-1}e^{-\frac{1}{2}\left( \kappa x+\frac{\chi}{x}\right)  }$, $x > 0$.
\item The posterior density of $\tau_k$ does not have a conjugate distribution. The conditional posterior sampling distribution of $\tau_k$ is given by
\begin{equation} \nonumber
\pi\left(\tau_k \mid \beta_{jk} \lambda_{j} \right) \propto \tau_k^{-p}exp\left[-\frac{1}{\tau_k^2}\sum_{j=1}^p \frac{\beta_{jk}^2}{2\lambda_{j}}\right]\frac{\gamma^2}{\left( \tau^2_k+\gamma^2\right)}. \label{tau density}
\end{equation}
For each $k=1,2, \cdots, K$, an adaptive Metropolis-Hastings (MH) step is applied to attempt an update to $\tau_k$, based on algorithm 4 of \cite{andrieu2008tutorial} applied to $\tau_k$.
\item Similarly, $c$ does not have a conjugate sampling density. The conditional posterior density of $c$ is given by
\begin{equation} \nonumber
\pi\left(c \mid \lambda_1,\lambda_2,\hdots,\lambda_p \right)  \propto  \frac{c^{cp}}{{\Gamma\left(c\right)}^p}exp\left[-c\left(\lambda_c+\sum_{j=1}^p \lambda_{j} \right) +(c-1) \sum_{j=1}^p log \lambda_j\right], \, c>0 . \label{c density}
\end{equation}
An adaptive MH step based on the \cite{andrieu2008tutorial} algorithm is also performed here.
\item $\Psi$ is drawn from $Inv-Wishart \left( \upsilon_0+n, S_0+S\right)$, where $S=\left( Y-XB \right)^T\left( Y-XB \right)$.
\end{enumerate}

\subsection{MOHS Model}
Sampling steps for MOHS model are described below.
\begin{enumerate}[label=(\roman*)]
\item Sampling distribution for $vec\left(B\right) \mid X, \Psi, \lambda, \tau$ is the same as in MONG step (i), except $\Lambda=diag(\lambda_1^2, \lambda_2^2, \cdots, \lambda_p^2)$ here.
\item For $j=1,2,\cdots,p$, sample $\lambda^2_j \mid \beta_{jk}, \tau_k,\nu_j    \sim   IG\left( \frac{K+1}{2}, \frac{1}{\nu_j}+\sum_{k=1}^K \frac{\beta_{jk}^2}{2\tau^2_k} \right).$
\item For $k=1,2,\cdots,K$, sample $\tau^2_k \mid \beta_{jk},\lambda_j,\epsilon_k \sim IG\left( \frac{p+1}{2}, \frac{1}{\omega_k}+\sum_{j=1}^p \frac{\beta_{jk}^2}{2\lambda^2_j} \right).$ 
\item For $j=1,2,\cdots,p$, sample $ \nu_j \mid \lambda_j \sim IG\left( 1, 1 + \frac{1}{\lambda^2_j} \right). $
\item For $k=1,2,\cdots,K$, sample $ \omega_k \mid \tau_k \sim IG\left( 1, 1 + \frac{1}{\tau^2_k} \right). $
\item Sample $\Psi \mid B \sim Inv-Wishart \left( \upsilon_0+n, S_0+S\right)$.
\end{enumerate}

\subsection{MODL Model}
For the original DL specification, \citep{Bhatt} propose a block sampler that involves marginalizations over different sets of parameters.  Due to sharing $\phi_j$s across multiple outcome models, this is no longer feasible in our MODL model $\left(\ref{Dir prior}\right)$, and we require (adaptive) Metropolis-Hasting to jointly sample the vector $(\phi_1,\ldots,\phi_p)$ of local parameters. Sampling steps are as follows:
\begin{enumerate}[label=(\roman*)]
\item First sample $vec(B)\mid X,\Psi,\phi,\eta,\tau$ from $MVN_{pK}\left(M ,W\right) $. The conditional posterior distribution of $vec(B)$ is as in the case of NG prior except $\Omega=diag\left(\eta_{jk}\phi_j^2 \tau_k^2\right)$.
\item For $k=1,2,\cdots,K$, sample $\tau_k \mid \beta_{jk}, \phi$ (marginalizing over $\eta$) from a generalized inverse Gaussian distribution $giG\left( pa-p,1,2\sum_{j=1}^{p}\frac{\beta_{jk}}{\phi_j}\right)$. 
\item The conditional posterior density of $\phi| B,\tau$ (marginalizing over $\eta$) is proportional to
\begin{equation}
\pi \left ( \phi_1,\phi_2,\hdots,\phi_p \mid  B,\tau \right) \propto \prod_{j=1}^p \phi_j ^{a-K-1} exp \left[ - \frac{1}{\phi_j } \sum_{k=1}^K \frac{\mid \beta_{jk} \mid}{\tau_k}  \right], \label{Phi_density} 
\end{equation}
where $\phi$ resides in the $\left(p-1\right)$-dimensional simplex. We have used an adaptive MH algorithm by extending algorithm 4 of \cite{andrieu2008tutorial} for sampling $\phi$. We sample from distribution $\left(\ref{Phi_density}\right)$ as described below: 
\begin{itemize}
\item At the $t^{th}$ iteration, sample the proposed move by 
\begin{equation}
\left(\phi_1^*,\phi_2^*,\hdots,\phi_{p}^* \right) \sim Dirichlet \left(\zeta^{\left(t\right)} \phi_1,\zeta^{\left(t\right)}  \phi_2,\hdots,\zeta^{\left(t\right)}  \phi_{p} \right). \label{proposal distn}
\end{equation}
The $\zeta^{\left(t\right)} $ is a positive tuning parameter that controls the dispersion of the proposal distribution. Note that this choice behaves similarly to a random walk with $E \phi_j^*= \phi_j$ and $Var \left(\phi_j^*\right)=\frac{\phi_j\left(1-\phi_j\right)}{1+\zeta^{\left(t\right)}}$. The variance of our candidate is inversely related to $\zeta^{\left(t\right)}$.
\item Calculate the MH probability $\alpha=min\left( 1,\frac{\pi\left( \phi^*|B,\tau \right) \, g\left(  \phi_1,\phi_2,\hdots,\phi_{p} \mid\phi_1^*,\phi_2^*,\hdots,\phi_{p}^* \right) }{\pi\left( \phi|B,\tau \right) \,  g\left( \phi_1^*,\phi_2^*,\hdots,\phi_{p}^*  \mid \phi_1,\phi_2,\hdots,\phi_{p} \right)}\right) $, where $g(\cdot)$ is the proposal distribution $\left(\ref{proposal distn}\right)$. With probability $\alpha$, we accept the proposed value $\phi^*=\left(\phi_1^*,\phi_2^*,\hdots,\phi_{p}^* \right)$, and otherwise, we retain the current $\phi=\left(\phi_1,\phi_2,\hdots,\phi_{p}\right)$.
\item Updating the tuning parameter $\zeta$ : 
\[log(\zeta^{\left( t+1 \right)})=log(\zeta^{\left(t\right)})-\gamma^{\left(t+1\right)} \left(\alpha-\alpha^*\right),  \]
where $\alpha^* =0.24$ is the ideal acceptance probability and the step size is $ \gamma^{\left(t\right)}=\min\left(500^{-\frac{1}{2}},t^{-\frac{1}{2}}\right)$. 
\end{itemize}
\item Sample $\eta_{jk}^{-1} \mid  \beta_{jk}, \phi, \tau$ independently from $Inv-Gaussian\left( 1,\frac{\phi_j \tau_k}{|\beta_{jk}|}\right)$. The Inverse Gaussian distribution is defined by the density function $f\left(x; \mu,\theta\right)= \left(\frac{\theta}{2 \pi x^3}\right)^{\frac{1}{2}} exp\left[ -\frac{\theta\left(x-\mu\right)^2}{2\mu^2x}\right],\, x > 0$. 
\item Sample $\Psi \mid B \sim Inv-Wishart \left( \upsilon_0+n, S_0+S\right)$.
\end{enumerate}

\section{Simulation Study}\label{section 4}

Here we implement simulation studies to evaluate the performance of our methodology. In addition to our MONG, MOHS, and MODL methods, we consider the following competitors: 
\begin{itemize}

\item \textbf{Naive Normal-gamma Model:} To assess the utility of sharing the local parameters across all response variables, we consider an approach that fails to make use of this information by independently placing a NG prior on the vector of regression coefficients $\left(\beta_{1k}, \beta_{2k}, \cdots, \beta_{pk} \right)$ for each model $k$. This naive model is unable to borrow strength across models to inform the shared level of sparsity. To that end, $\beta_{jk} \sim N\left( 0, \lambda_{jk} \tau_k^2\right) $, where all $\lambda_{jk}$ are independent from $Gamma\left(c,\frac{1}{c}\right)$. The rest of the model is unaffected.

\item \textbf{Naive Horseshoe Model:} Similar to the naive NG model, we consider applying a horseshoe prior independently for each response. In this case, $\beta_{jk} \sim N\left( 0, \lambda_{jk}^2 \tau_k^2\right) $, with all $\lambda_{jk}$ independently from $C^+(1)$.

\item \textbf{Naive Dirichlet-Laplace Model:} We also consider a naive version of DL prior. To that end, we let $\beta_{jk} \sim DE\left( \phi_{jk} \tau_k\right) $. Here, independent local shrinkage parameters are drawn for each response model $k$: $\phi_{k}=\left( \phi_{1k}, \phi_{2k}, \cdots, \phi_{pk}\right) \sim Dir(a,a, \ldots, a)$.

\item \textbf{No Shrinkage Model:} As a baseline that does not favor any variable selection, we consider a basic conjugate prior model. For all $j, k$, $\beta_{jk} \sim N\left( 0,10 \right) $ to provide minimal shrinkage towards zero.

\item \textbf{Selection Prior \citep{brown1998multivariate} Model:} As noted in the introduction, this approach constrains each predictor to either be in the model of all $K$ responses or to be excluded from all.

\item \textbf{MBSP Model \citep{bai2018high}:} As previously noted, this approach is similar to our MOHS model where the global parameter $\tau$ is common across all responses and fixed by asymptotic arguments, rather than estimated from the data. The performance of this model is obtained using their available R package MBSP. 
\end{itemize}

Data are generated from the multi-response linear regression model $\left(\ref{multioutcome_model}\right)$ using a design matrix $X^{n \times p}$ whose elements are independently drawn from a standard normal distribution. Then, rows of the response matrix $Y^{n \times p}$ are independently generated from $N_K\left(X_{i_{\cdot}}B,\Psi\right)$, where $\Psi_{ij}=0.5$ if $i \neq j$, and 1 otherwise. We consider $p=20$ predictors, $K=10$ response variables, and a sample size of $n=500$. We generate 100 datasets, and for each dataset and model choice we run the MCMC chain for 90,000 iterations with a burn-in of 10,000 iterations. We measure predictive performance by computing the mean square prediction error (MSPE) using the posterior mean regression coefficients $\hat{B}$ and an independently generated test data set. To assess the accuracy of the regression coefficient estimation, we consider the sum of square errors (SSE). To distinguish between error of over-shrinking relevant signals and under-shrinking non-signals, we partition this SSE into the SSE over the true non-zero $\beta_{jk}$s and the SSE for the $\beta_{jk}$s that are true zeros. These quantities are determined by the following formulas:
\begin{equation} \nonumber
MSPE= \frac{1}{Kn_{test}}\sum^{K}_{k=1}\sum^{n_{test}}_{i=1} \left(X_{i\cdot}^{\left(t\right)}\hat{B_{\cdot k}}-Y_{ik}^{\left(t\right)} \right)^2  \quad \quad
SSE= \sum^{K}_{k=1}\sum^{p}_{j=1} \left(\hat{\beta}_{jk}-\beta_{jk} \right)^2,
\end{equation}
where $n_{test}$ is the number of observations in the test dataset $\left(n_{test}=500\right)$, and $Y^{\left(t\right)}$ and $X^{\left(t\right)}$ denote respectively the response and design matrices for the test set. We consider two scenarios for choosing the true regression coefficient structure. First, we consider a simple sparse $B^{\left( 0\right) }$ matrix (Table \ref{beta matrix}), where each covariate is either important for all responses or has no contribution to the mean of any response.   Table \ref{beta table} presents the results for this case.



\begin{small}
\begin{table}[H]
\centering
\caption{True $B^{(0)}$ regression coefficient matrix in first simulation study.}
\label{beta matrix}
\[
B^{(0)}
=
\begin{pmatrix}
    2.0 & 2.0 & 2.0 & 2.0 & 2.0 & 2.0 & 2.0  & 2.0 & 2.0 & 2.0 \\
   -3.0 & -3.0 &-3.0 & -3.0 & -3.0 & -3.0 & -3.0  & -3.0 & -3.0 & -3.0 \\
    1.0 & 1.0 & 1.0 & 1.0 & 1.0 & 1.0 & 1.0  & 1.0 & 1.0  & 1.0 \\
    0.0 &  0.0  & 0.0 & 0.0 &0.0 &0.0 &  0.0   &  0.0  &  0.0   &  0.0 \\
     
    \vdots & \vdots & \vdots & \vdots & \vdots & \vdots & \vdots & \vdots & \vdots & \vdots \\
    0.3 & 0.3 &0.3 & 0.3  & 0.3 & 0.3 & 0.3 & 0.3 & 0.3 & 0.3\\
     0.0  &  0.0  & 0.0  &  0.0 & 0.0  &  0.0 &  0.0   &  0.0 & 0.0  &  0.0   \\
     0.0  &  0.0  & 0.0  &  0.0 & 0.0  &  0.0 &  0.0   &  0.0 & 0.0  &  0.0    \\
\end{pmatrix}
\]
\end{table}
\end{small}

Comparing each of our multi-outcome models to their respective naive versions, we find reduced MSPE in all cases.  While the difference in MSPE between models are relatively minor, there are large improvements in the coefficient estimation. Our shared shrinkage models lead to reduction in total SSE of around $50\%$ when compared to the respective naive version. When looking at the two components of SSE, we see clear improvement in the estimation of the coefficients that are truly zero. That is, by sharing the local parameters across the $K$ outcome models, our model is able to better identify those coefficients that should be aggressively shrunk toward zero. Our proposed model also yields similar level of predictive performance with the selection prior approach \citep{brown1998multivariate}, which is perfectly suited to this choice of $B^{\left(0\right)}$.

\begin{small}
\begin{table}[H]
\centering
\caption{Prediction and estimation results from simulation study with $B^{(0)}$. }
\label{beta table}
\begin{tabular}{lcccc}
\hline \hline
\multirow{3}{*}{Models}                  
                                         & \multirow{2}{*}{MSPE} & \multicolumn{3}{c}{SSE}                 \\ \cline{3-5}  
                                         &                       & All $\beta$ & $\beta \neq$0 & $\beta$=0  \\ \hline \hline
MONG      & 1.028  & 0.097 & 0.095        & 0.002      \\
MODL$\left(a=0.5\right)$    & 1.029   & 0.098     & 0.089        & 0.009      \\
MODL$\left(a=1/p\right)$   & 1.032     & 0.125      & 0.112        & 0.013    \\
MOHS           & 1.030    & 0.113      & 0.099        & 0.014     \\
Naive NG       & 1.040     & 0.205       & 0.167         & 0.038  \\
Naive DL$\left(a=0.5\right)$  & 1.036   & 0.176  & 0.104  & 0.073   \\
Naive DL$\left(a=1/p\right)$   & 1.079  & 0.564  & 0.547 & 0.016  \\
Naive Horseshoe     & 1.040  & 0.203  & 0.111  & 0.092   \\
No shrinkage       & 1.059   & 0.416  & 0.080  & 0.337   \\
Selection prior  & 1.028 & 0.134 & 0.134 & 0.000 \\
MBSP model  & 1.029  & 0.104  & 0.092 & 0.012\\
\hline \hline
\end{tabular}
\end{table}
\end{small}

We note that the model without shrinkage is not competitive due to its large SSE in the zero coefficients. Also the naive DL with $a=\frac{1}{p}$ performs poorly in estimating the non-zero coefficients.  Setting $a=\frac{1}{p}$ provides a much stronger level of shrinkage than the $a=0.5$ case.  For the naive DL model, we do see more shrinkage under $a=\frac{1}{p}$ than $a=0.5$, but by sharing shrinkage information across multiple responses, our MODL model is able to find an acceptable balance in the amount of shrinkage under both choices of $a$.



\begin{small}
\begin{table}[]
\centering
\caption{True $B^{(1)}$ regression coefficient matrix in second simulation study.}
\label{beta_perm matrix}
\[
B^{(1)}
=
\begin{pmatrix}
    2.0 & 2.0 & 2.0 & 2.0 & 2.0 & 2.0 & 2.0  & 2.0 & 2.0 & 2.0 \\
     -3.0 & -3.0 &-3.0 & -3.0 & -3.0 & -3.0 & -3.0  & -3.0 & -3.0 & -3.0 \\
    1.0 & 1.0 & \textbf{0} & 1.0 & \textbf{0} & 1.0 & 1.0  & 1.0 & 1.0  & 1.0 \\
    0.0 &  0.0  & \textbf{0.5} & 0.0 &0.0 &0.0 &  0.0   &  0.0  &  0.0   &  0.0 \\
     0.0  &  0.0 & 0.0 & 0.0 &0.0 & 0.0 &  \textbf{0.3  } &  0.0  &  0.0   &  0.0 \\
    \vdots & \vdots & \vdots & \vdots & \vdots & \vdots & \vdots & \vdots & \vdots & \vdots \\
    0.0  &  0.0 & 0.0 & 0.0 & 0.0 & 0.0 &  \textbf{1.5  } &  0.0  &  0.0   &  0.0 \\
     \vdots & \vdots & \vdots & \vdots & \vdots & \vdots & \vdots & \vdots & \vdots & \vdots\\
     0.0  &  0.0  & 0.0  &  0.0 &0.0  &  0.0 &  0.0   &  0.0 &  0.0   &  0.0   \\
    0.3 & 0.3 &0.3 & 0.3  & \textbf{0} & 0.3 & 0.3 & 0.3 & 0.3 & 0.3\\
     0.0  &  0.0  & 0.0  &  0.0 & 0.0  &  0.0 &  0.0   &  0.0 & 0.0  &  0.0   \\
     0.0  &  0.0  & 0.0  &  0.0 & 0.0  &  0.0 &  0.0   &  0.0 & 0.0  &  0.0    \\
\end{pmatrix}
\]
\end{table}
\end{small}

Next, we consider a situation that does not have the exact same sparse structure for each response model.  There are two important considerations for such a choice.  First, in light of our original motivation, we are interested in a more flexible model than those  require the same subset of predictors for all responses. We wish to assess the performance of our model in such a case where there are variations in the relevant predictors across models. An alternative motivation is to understand the impact of misspecification for models that assume the exact same subset of relevant predictors across all outcomes.  To that end, the new true coefficient matrix $B^{(1)}$ in Table \ref{beta_perm matrix} is created by perturbing $B^{(0)}$ so that the true model no longer has exact sparsity across all models. We switch three of the zero coefficients from $B^{(0)}$ to non-zero $\beta_{jk}$ and also change three non-zero coefficients in $B^{(0)}$ to zero (as denoted in bold). This potentially represents a more realistic scenario where a small subset of predictors impact all responses, but there are some minor deviations from this general rule.

\begin{small}
\begin{table}[]
\centering
\caption{Prediction and estimation results from simulation study with $B^{(1)}$. }
\label{beta perm table}
\begin{tabular}{lcccc}
\hline \hline
\multirow{3}{*}{Models}             
                                    & \multirow{2}{*}{MSPE} & \multicolumn{3}{c}{SSE}  \\ \cline{3-5}  
 &   & All $\beta$  & $\beta \neq 0$    & $\beta=0$     \\ \hline \hline
MONG          & 0.976     & 0.113  & 0.095  & 0.028  \\
MODL$\left(a=0.5\right)$    & 0.976    & 0.118  & 0.085  & 0.033   \\
MODL$\left(a=1/p\right)$  & 0.978    & 0.127  & 0.097  & 0.030  \\
MOHS        & 0.978     & 0.132  & 0.098  & 0.034  \\
Naive NG     & 0.978   & 0.131  & 0.091  & 0.040  \\
Naive DL$\left(a=0.5\right)$    & 0.980   & 0.158  & 0.083  & 0.075   \\
Naive DL$\left(a=1/p\right)$            & 0.994   & 0.283  & 0.251  & 0.032  \\
Naive Horseshoe            & 0.982   & 0.177   & 0.083   & 0.094      \\
No shrinkage                   & 1.005   & 0.416  & 0.080  & 0.336 \\
Selection prior  & 0.979 & 0.139 &  0.090 & 0.050 \\
MBSP model  & 0.978 & 0.146  & 0.080 & 0.066   \\
\hline \hline
\end{tabular}
\end{table}
\end{small}

The results for this simulation settings are reported in the Table \ref{beta perm table} and are generally similar to the previous analysis.  As would be expected, the gap between the shared shrinkage and the naive approaches is somewhat narrowed, but the proposed approaches continue to show lower MSPE and lower SSE than their naive counterparts in all cases.  Hence, even if there are some differences in which predictors are relevant across models, sharing shrinkage information through our common local parameter structure can continue to improve estimation. The selection prior approach and MBSP model also show similar prediction performance, although both have poorer performance in the coefficient estimation  relative to our approach. Of particular note, the MBSP has fairly large SSE for the zero signals, indicating a lower level of shrinkage than our proposals. Our model estimate the global parameters from the data to adjust the amount of shrinkage, whereas MBSP fixes $\tau$ and is unable to correct for undershrinkage in this data.

In conclusion, our three multi-outcome models perform well in those simulation studies. Using $a=\frac{1}{p}$ in the MODL model may lead to overshrinking, so we typically prefer $a=0.5$. While the differences between methods are relatively minor, MONG tends to perform best among our proposals.

\section{Application}\label{section 5}

We now demonstrate our methodology with the yeast cell cycle data set \citep{spellman1998comprehensive} from the spls package in R. The data was first analyzed by \cite{chun2010sparse} and also by \cite{bai2018high}. In this dataset, the response matrix $Y$ contains gene expression data for $n=542$ genes from an $\alpha$ factor based experiment. Each column of $Y$ corresponds to mRNA levels measured at 7 minute intervals across 2 hours providing a total of $K=18$ responses. The covariate matrix $X$ contains the binding information for $p=106$ transcription factors (TFs).  In molecular biology, transcription factors are a diverse family of proteins which are involved in the process of transcribing, DNA into RNA. Hence, it is of common interest to identify the most significant TFs that play an important role in gene regulations. 

We applied our method to capture those TFs that affect the expression levels across all time points. We perform the analysis using our proposed MONG, MODL, and MOHS models, followed by the three naive models, the no shrinkage model, the selection model \citep{brown1998multivariate} and the MBSP model \citep{bai2018high}. Due to over-shrinkage observed in the MODL$\left(a=\frac{1}{p}\right)$ model, we do not consider its performance here. For each case, we run a burn-in for 1000 iterations followed by another 30,000 iterations. We report the MSPE by performing cross validation on 50 data sets for each model to assess the predictive power of each method. For cross validation we randomly assign $80\%$ of observations to the training set to estimate $B$, and then measured the MSPE using the remaining $20\%$. We also analyze the full dataset and compute the deviance information criteria $DIC$ as a model comparison measure \citep{spiegelhalter2002bayesian}. $DIC$ is calculated by $DIC=D+2p_{D}$, where $D$ is the deviance at the posterior expectation of the parameter values and $p_{D}$ is the effective number of parameters, and smaller $DIC$s are favored. $p_D$ is calculated as $p_D= E\left\lbrace D\left(B,\Psi|Y\right)\right\rbrace-D(\hat{\Psi},\hat{B})$. Table \ref{Yeast table} shows the MSPE, the deviance at the posterior expectation of the parameter values $\left(D\right)$, the effective number of parameters $\left(p_{D}\right)$, and the deviance information criteria (DIC) of the yeast cycle data for each model.


\begin{small}
\begin{table}[H]
\centering
\caption{Cross-validation prediction error and model comparison statistics for yeast cell cycle data.}
\label{Yeast table}
\begin{tabular}{lcccc}
\hline \hline
\multicolumn{1}{c}{Model} & MSPE   & $D$ & $p_{D}$      & DIC          \\ \hline \hline
MONG                      & 0.833 & 15580  & 370  & 16321  \\
MODL$\left(a=0.5\right)$               & 0.814 & 14077 & 1299 & 16676  \\
MOHS                      & 0.841 & 15683 & 318  & 16320  \\
Naive NG                  & 0.987 & 16594  & 148  & 16890  \\
Naive DL$\left(a=0.5\right)$             & 0.907 & 13990  & 1430 & 16851 \\
Naive DL$\left(a=1/p\right)$ & 0.872 & 15117	& 733 &	16584\\
Naive HS                  & 0.864 & 14706  & 827 & 16361 \\
No shrinkage             & 0.971 & 13453 & 2131 & 17716 \\
Selection prior            & 0.845 & 17425 & 257 & 17940   \\ 
\hline \hline
\end{tabular}
\end{table}
\end{small}

The MODL$\left(a=0.5\right)$ choice yields the lowest prediction error among our models.  Consistent with the simulation study, each of the multi-outcome approaches have smaller MSPE than their respective naive counterparts. The MONG and MOHS model also yield a lower mean square prediction error by slightly outperforming the selection prior model. 

When using the competitor MBSP model, the prediction error is 0.786, scoring lowest among all approaches. It appears that for this particular data application, using a fixed value of $\tau$ performs slightly better than our methods which require estimating $K=18$ global parameters. However, as  noted in the simulation study, this is not always the case, and worse performance may result. Finally, we note that the R package of MBSP model only produces model estimates and not the full set of posterior samples. So we were unable to compute DIC estimates for the MBSP model.

The DIC criteria favors the MONG and MOHS models.  When considering the effective number of parameters, we see that these models estimate a much sparser regression coefficient matrix than MODL.  When comparing DIC between the shared shrinkage and naive models, we again see that our proposals consistently dominate their counterparts that fail to share variable selection information between responses. The selection approach from \cite{brown1998multivariate}, which requires a common set of predictors for all models performs poorly with respect to DIC. This model places the majority of the posterior probability on models with only 2 or 3 predictors. This excessive sparsity leads to high prediction error, poor model fit, and large DIC.

\begin{figure}[H]
\centering
    \includegraphics[scale=0.3]{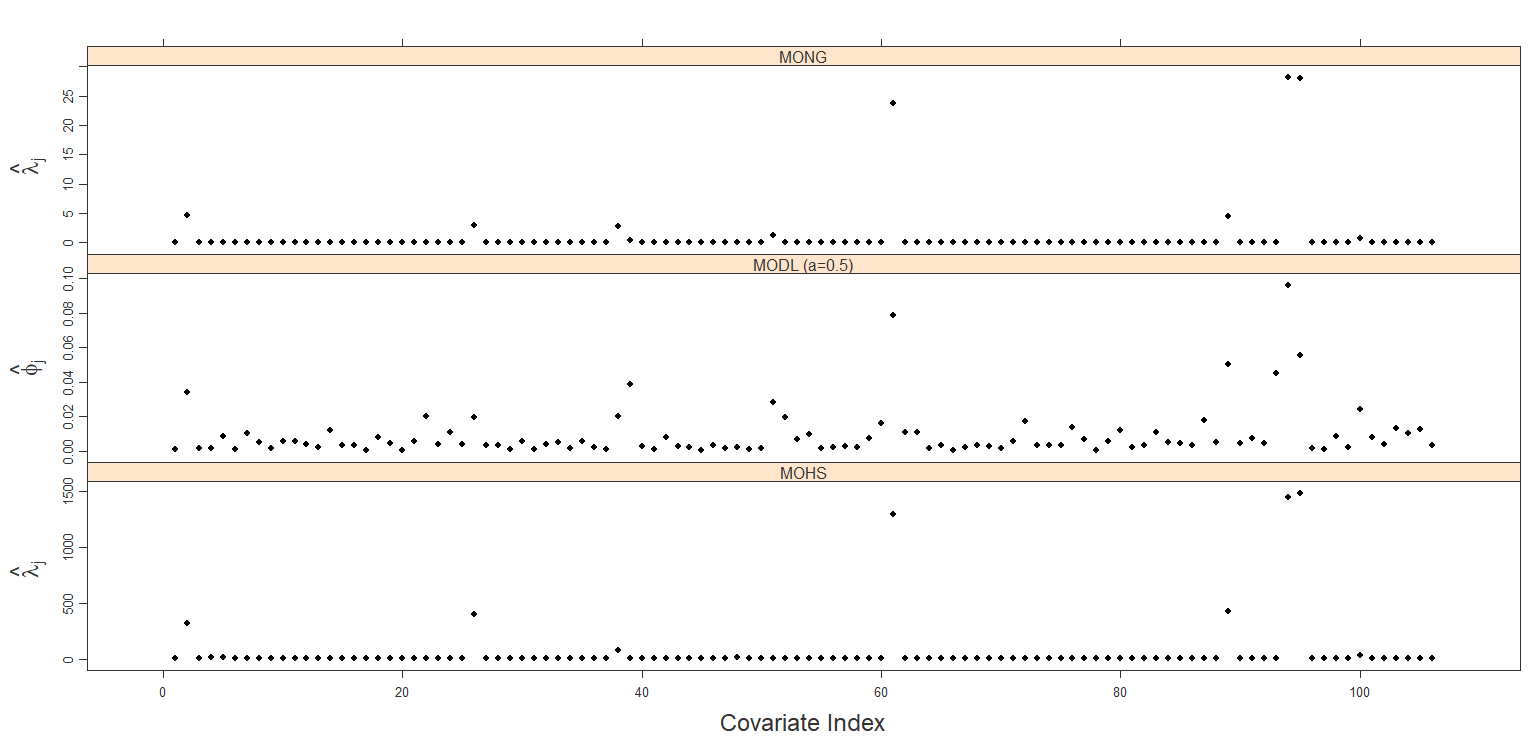}
\caption{Estimated local parameter $\left(\hat{\lambda}_j \,\, \text{or} \,\, \hat{\phi}_j\right)$ across all predictors in the three multi-outcome regression analyses for the yeast cell data.}
    \label{Yeast_lambda}
\end{figure}

Based on the results from fitting the full data set, we consider the use of the local parameters as a marker of variable importance. Figure 1 graphically displays these parameters for each of  the multi-outcome models. Based on the MONG results, we would consider those covariates with $\hat{\lambda}_j>1$ as evidence of a strong effect across all response models. This criterion selects 8 important TFs: SWI5, SWI6, NDD1, ACE2, STE12, HIR1, GAT3, MBP1. The 8 predictors with the largest $\hat{\lambda}_j$ in the MOHS model corresponds to the same 8 TFs, indicating robustness in the predictor weights across the model variations. Consistent with its large $p_D$ indicating less sparsity, the MODL choice demonstrates much less separation between large and small $\phi_j$ and consequently less shrinkage/sparsity in the $\hat{B}$ matrix. For this MODL case, distinguishing important predictors based on the value of the local parameters will not be effective.



\section{Discussion}\label{section 6}
In this paper, we have proposed a general strategy of variable selection in the multivariate regression model by sharing common local parameters across all of the response variables.  We have demonstrated our approach using the Normal-gamma, Dirichlet-Laplace and horseshoe priors.  Based on the results from simulation studies and the analysis of data from an mRNA experiment, we have demonstrated the utility of our approach in comparison to alternatives.  Our approaches are found to be superior in terms of both predictive performance and parameter estimation.  In general, we recommend the use of the MONG version of our model as it displayed consistently strong behavior across all empirical experiments, although the MODL and MOHS also performed well.

Regarding computational comparisons between our methods, the MOHS model tends to run fastest as all of its sampling distributions are conditionally conjugate.  While slightly slower, MONG has comparable computational time for a fixed number of iterations.  However, the MODL model tends to be computationally slower due to the sampling of $pK$ data augmentation parameters $\eta_{jk}$.  Moreover, as noted in Section 3, the mixing in this algorithm tends to be slower due to the multivariate MH sampling of $\phi=(\phi_1,\ldots,\phi_p)$.  While our adaptive step is generally effective here, further algorithmic improvements may be possible here in future research.




\vspace{5mm} 

\noindent Jeremy T. Gaskins, Department of Bioinformatics and Biostatistics, University of Louisville, 485 E Gray St., Louisville, 40202, USA.\\
Email: jeremy.gaskins@louisville.edu

\appendix

\renewcommand{\theequation}{A.\arabic{equation}}
\setcounter{equation}{0}

\setcounter{secnumdepth}{0}
\section{Appendix}

Here we provide details and the proof of the posterior consistency results from Section 2.5.

For our discussion we use the term multi-index to denote a model where the individual observations belong to a common multidimensional family $f(\cdot)$ but are indexed by possibly different parameters $\theta_{iB}$. The second subscript denotes a global  parameter $B$, which in our context is the (shared) matrix of regression coefficients. Thus, in our multivariate Gaussian regression we let $\theta_{iB}= X_{i\cdot} B$ be the $K$-vector representing the mean of the $K$ responses. 

Recall that the KL distance between two densities is defined as $E_{f_0}\left\{\log \frac{f_1(Y)}{f_0(Y)}\right\}$. For multi-index families, we extend the definition to have a  notion KL distance for each $i$. To that end, the KL distance between the global parameter $B$ and the true value $B_0$ for observation $i$ can be written as 
\begin{equation} \nonumber
KL_i(B, B_0)=  E_{B_0} \left\{\log \left(\frac{f(Y_i, \theta_{iB})}{f(Y_i; \theta_{iB_0})}\right)\right\}, \label{ratio}
\end{equation}
where $\theta_{iB}= X_{i\cdot}B$ and  $\theta_{iB_0}= X_{i\cdot}B_0$ are parameter vectors indexing the densities for observation $i$ under parameters $B$ and $B_0$. We also define $V_i(B,B_0)$ as the variance analogue, i.e.,
\begin{equation} \nonumber
V_i(B, B_0)=  Var_{B_0} \left\{\log \left(\frac{f(Y_i, \theta_{iB})}{f(Y_i; \theta_{iB_0})}\right)\right\}. \label{KLvar}
\end{equation}
 
We first state our Lemma 1 which establishes a uniformly exponentially consistent (UEC) sequence of tests  that will be required in the proof of Theorem 1.  Here, we include the dependence on $n$ by letting $Y_n$ and $X_n$ denote the response and design matrices for a sample of size $n$.
\begin{lemma} \label{lem 1}
  For any $\epsilon>0$, define $B_\epsilon=\left\lbrace B :||{B-B_0}|| > \epsilon \right\rbrace$.  Let $\Phi_n=I\left(Y_n \in C_n\right)$ be the test statistic based on the critical region $C_n=\left\lbrace Y_n :||{\hat{B}-B_0}|| > \frac{\epsilon}{2} \right\rbrace$ and $\hat{B}=\left(X_n^TX_n\right)^{-1}X_n^TY_n$.  Further, assume condition (A3), and let $d$ be the largest eigenvalue of $\Psi$.  Then, for the likelihood $\left(\ref{multioutcome_model}\right)$, we have the following:
\begin{enumerate}
\item $E_{B_0}\left(\Phi_n\right) \leq \exp\left(-n\frac{\epsilon^2c}{16d}\right)$,
\item $\sup_{B \in B_\epsilon} E_{B}
\left( 1-\Phi_n \right) \leq \exp\left(-n\frac{\epsilon^2c}{16d}\right).$
\end{enumerate}
\end{lemma}  
\noindent {\textit{Proof}:} Proof of this lemma follows as in Lemma 1 of \cite{bai2018high}. 

Next, we state and prove  Lemma 2 which establishes the prior positivity condition.

\begin{lemma} \label{lem 2}
 Assume a fixed $\Psi$, the likelihood $\left(\ref{multioutcome_model}\right)$, (A1), and (A2). Then, for all $\epsilon>0$, there exists a set $C_\epsilon$ with $\pi(B\in C_{\epsilon})>0$, such that for all $B \in C_{\epsilon}$
\begin{eqnarray} \nonumber
   KL_i(B, B_0) &<& \epsilon \hspace{1em}\text{for all $i$}, \\
\sum_{i=1}^{\infty} \frac{1}{i^2}V_i(B, B_0 ) &<& \infty.   \nonumber   
\end{eqnarray}
\end{lemma}  
\noindent { \textit{Proof}:}
A little algebra shows that
\[ KL_i(B,B_0) = ( {X}_i B- {X}_i B_0) \Psi^{-1}  ( {X}_i B- {X}_i B_0)^{\prime}. \]
Let $\tilde{X}_i = I_K \otimes X_{i\cdot}$, $\beta=\text{vec}(B)$, and $\beta_0=\text{vec}(B_0)$.  Then, it follows that 
\begin{eqnarray*}
KL_i(B,B_0) &=& ( {X}_i B- {X}_i B_0)  \Psi^{-1}  ( {X}_i B- {X}_i B_0)^{\prime} = ( \tilde{X}_i \beta- \tilde{X}_i \beta_0)^{\prime}  \Psi^{-1}  ( \tilde{X}_i \beta- \tilde{X}_i \beta_0) \\
&=&   (\beta-\beta_0) \tilde{X}_i \Psi^{-1} \tilde{X}_i^{\prime} (\beta-\beta_0)^{\prime} =
|| M_i (\beta-\beta_0)||^2, 
\end{eqnarray*}
where $M_i= \Psi^{-\frac{1}{2}}\tilde{X}_{i\cdot}$\,.
From the sub-multiplicativity of the Frobenius norm, $||M_i||$ is bounded by $||\Psi^{-\frac{1}{2}} ||\,||\tilde{X}_{i}||= K^{1/2}||\Psi^{-\frac{1}{2}} ||\,||{X}_{i\cdot}||$, which is bounded by $GK^{1/2}||\Psi^{-\frac{1}{2}} ||$ using (A2).  Clearly, $||\beta-\beta_0||=|| B-B_0||$.  Thus, a set  $C_{\epsilon}= \left\lbrace B: ||B-B_0|| < \frac{\epsilon}{ GK^{1/2}||\Psi^{-\frac{1}{2}} ||} \right\rbrace$  will clearly satisfy $KL_i(B, B_0) < \epsilon $ for all $i$. By (A1) the continuous prior $\pi(B)$ assigns positive probability to any such open neighborhood $C_{\epsilon}$. Similar steps show that for all $B$ in $C_\epsilon$, the $V_i$s are bounded uniformly by a constant across all $n$, proving convergence of $\sum_{i=1}^{\infty} \frac{1}{i^2}V_i(B, B_0 )$.

We first introduce and sketch the proof of a more general theorem that establishes posterior consistency for a wide range of multi-index models.

 \begin{theorem} 
Consider a multi-index model with global parameter $B$ and independent observations $Y_i$, $i=1, \ldots, n,\ldots$ with $Y_i\sim f(\cdot, \theta_{iB_0})$ under the true global parameter value $B_0$.  Further assume the following two conditions:
  \begin{enumerate} 
  \item There exist tests $\Phi_n$  such that  $ E_{B_0} (\Phi_n) < exp(-nC_1)$ and that for all $B\ne B_0$, $ E_{B}(1-\Phi_n) < exp(-nC_2 )$. Here, $C_1$ and $C_2$ are constants not depending on the parameter of interest.  \label{condition 1}
\item   There exists a set $C_{\epsilon}$ with $\pi(B\in C_{\epsilon})>0$, such that for all $B \in C_{\epsilon}$, \label{condition 2}
\begin{eqnarray} \nonumber
   KL_i(B, B_0) & <  &\epsilon \hspace{1em}\text{for all $i$}, \\
\sum_{i=1}^{\infty} \frac{1}{i^2}V_i(B, B_0 ) & < &\infty  \nonumber 
\end{eqnarray}

\end{enumerate} 
Then, the posterior distribution for $B$ is consistent.  That is, for any $\epsilon>0$,
\[P_{B_0}\left\{ \,||B- B_0||>\epsilon \mid Y_{1},\ldots,Y_{n}\right\} \rightarrow 0, \hspace{1em} \text{a.s. as $n\rightarrow \infty$.}\]  
\end{theorem}

\noindent {\textit{Proof}:}
The proof of this theorem is a combination of arguments in \cite{scr}, \cite{ame} and \cite{choir}, and we omit the technical details.  Briefly the argument is as follows.  The posterior probability of interest, denoted by $L^\epsilon_n$, can be written as a ratio  of integrals of two likelihood ratios in the following way
\begin{equation} \nonumber
 L^\epsilon_n = P_{B_0}\left\{\, ||B- B_0||>\epsilon \mid Y_{1},\ldots,Y_{n}\right\} =\frac{ \int_{U_{\epsilon}}  \frac{\prod_i f(Y_i, \theta_{iB}) }{\prod_i f(Y_i, \theta_{iB_0}) } dB } {                           \int_{U}  \frac{\prod_i f(Y_i, \theta_{iB})}{\prod_i (f(Y_i, \theta_{iB_0}) } dB },
\label{rat}
\end{equation} 
   where  $U_\epsilon= \{B: ||B- B_0|| > \epsilon\}$ is the $\epsilon$-ball around $B_0$ and  $U$ is the entire parameter space. The aim is to show $L^{\epsilon}_n$ converges to 0 a.s.\ under $P_{B_0}$ for all $\epsilon>0$.

  As shown in  \cite{scr}, we may bound $L^\epsilon_n$ using the test statistic $\Phi_n$   as 
  \[  L^\epsilon_n \leq \Phi_n + \frac{  J_{1n} } {J_{2n}}, \]
where $J_{1n}=  \int_{U_{\epsilon}}  \frac{ (1- \Phi_n) \prod_i f(Y_i, \theta_{B_i})}{\prod_i f(Y_i, \theta_{B_0}) } dB$ and $J_{2n}=\int_{U}  \frac{\prod_i f(Y_i, \theta_{B_i})}{\prod_i f(Y_i, \theta_{B_0}) } dB$.
Following the arguments from \cite{scr} (also used in \cite{bai2018high} and \cite{armr}), the first condition in Theorem 2 can be shown to imply $\Phi_n \rightarrow 0$ a.s.  Further, $e^{nC} J_{1n} \rightarrow 0$ a.s., for a constant $C>0$ that may depend on auxiliary parameters (such as $\Psi$ and the eigenvalues of the design matrix) but not on $B_0$.  Similarly, the second condition of the Theorem 2 can be shown to imply that for any constant $c>0$, $e^{nc} J_{2n} \rightarrow \infty$ a.s.  In combination, these imply that $L^\epsilon_n$ converges almost surely to zero under the true parameter $B_0$, guaranteeing posterior consistency.

We note that the proof of this  theorem has a general flavor in that it only requires a UEC sequence of tests and prior positivity. The first condition can be satisfied for several settings involving multivariate Gaussian likelihoods. The second condition is applicable to a variety of model specifications and holds simply when observations are independent but not identically distributed. Of note, condition 2 was proved in \cite{scr} for single-index families and later adapted to multi-index families \citep{choir}. For a  proof of this, we refer  the  reader to  the proof of part A.5 in Theorem 1 from \cite{choir}.

\noindent { \textit{Proof of Theorem 1:}}
Results from Lemmas 1 and 2 are immediately obtained from assumptions (A1)--(A3), and these lemmas establish the two conditions required for Theorem 2.  Hence, Theorem 1 is proved.

\pagebreak

\end{document}